\documentclass[aps, prd, twocolumn, superscriptaddress, nofootinbib,
amsmath, amssymb]{revtex4-1}

\usepackage{graphicx}
\usepackage{mathrsfs} 
\usepackage[normalem]{ulem} 

\begin{document}

\title{Reconfinement, localization and thermal monopoles in $SU(3)$ trace-deformed Yang-Mills theory}

\author{Claudio Bonati}
\email{claudio.bonati@unipi.it}
\affiliation{Dipartimento di Fisica dell'Universit\`a di Pisa and INFN
  - Sezione di Pisa, Largo Pontecorvo 3, I-56127 Pisa, Italy.}

\author{Marco Cardinali}
\email{marco.cardinali@pi.infn.it}
\affiliation{Dipartimento di Fisica dell'Universit\`a di Pisa and INFN
  - Sezione di Pisa, Largo Pontecorvo 3, I-56127 Pisa, Italy.}

\author{Massimo D'Elia}
\email{massimo.delia@unipi.it}
\affiliation{Dipartimento di Fisica dell'Universit\`a di Pisa and INFN
  - Sezione di Pisa, Largo Pontecorvo 3, I-56127 Pisa, Italy.}

\author{Matteo Giordano}
\email{giordano@bodri.elte.hu}
\affiliation{ELTE E\"otv\"os Lor\'and University, Institute for
  Theoretical Physics, P\'azm\'any P.\ s.\ 1/A, H-1117, Budapest,
  Hungary.}

\author{Fabrizio Mazziotti}
\email{fabrizio.mazziotti@phd.unipi.it}
\affiliation{Dipartimento di Ingegneria dell'Informazione, Universit\`a di Pisa, 
Via Caruso 16, I-56122 Pisa, Italy.}

\begin{abstract}
  We study, by means of numerical lattice simulations, the properties
  of the reconfinement phase transition taking place in trace deformed
  $SU(3)$ Yang-Mills theory defined on $\mathbb{R}^3\times S^1$, in
  which center symmetry is recovered even for small compactification
  radii. We show, by means of a finite size scaling analysis, that the
  reconfinement phase transition is first-order, like the usual
  $SU(3)$ thermal phase transition. We then investigate two different
  physical phenomena, which are known to characterize the standard
  confinement/deconfinement phase transition, namely the condensation
  of thermal magnetic monopoles and the change in the localization
  properties of the eigenmodes of the Dirac operator. Regarding the
  latter, we show that the mobility edge signalling the Anderson-like
  transition in the Dirac spectrum vanishes as one enters the
  reconfined phase, as it happens in the standard confined
  phase. Thermal monopoles, instead, show a peculiar behavior: their
  density decreases going through reconfinement, at odds with the
  standard thermal theory; nonetheless, they condense at
  reconfinement, like at the usual confinement transition. The
  coincidence of monopole condensation and Dirac mode delocalization,
  even in a framework different from that of the standard confinement
  transition, suggests the existence of a strict link between them.
\end{abstract}

\maketitle

\section{Introduction}  

Yang-Mills (YM) theories defined on the manifold
$\mathbb{R}^3\times S^1$, where one of the directions is compactified,
undergo a phase transition as soon as the length $L_c$ of the
compactified direction becomes smaller than a critical length.  If the
compactified dimension is interpreted as the Euclidean time direction,
then the length $L_c$ is just the inverse of the temperature $T$ of
the system, and the phase transition is the well known finite
temperature deconfinement phase transition \cite{Cabibbo:1975ig,
  Polyakov:1978vu, Susskind:1979up, McLerran:1981pb}.

The deconfinement phase transition is associated with the spontaneous
breaking of center symmetry \cite{McLerran:1981pb, Svetitsky:1982gs,
  Yaffe:1982qf}, i.e., the invariance of the compactified theory under
gauge transformations which are periodic up to an element of the gauge
group center along the compactified direction. The order parameter
that signals the spontaneous breaking of center symmetry is the
Polyakov loop, i.e., the holonomy of the gauge field along the
compactified direction:
\begin{equation}
P(\vec{x}) = \mathcal{P}\mathrm{exp}\left( i g \int _0 ^{L_c}
  A_4(\vec{x}, \tau) \mathrm{d}\tau \right)\ , 
\end{equation}
where $\mathcal{P}$ denotes path-ordering.  Indeed, it is well known
that $\mathrm{Tr}P$ transforms non-trivially under center
transformations, and a nonzero value of $\langle \mathrm{Tr}P\rangle$
signals the finiteness of the free energy of an isolated static color
charge, i.e., deconfinement.

The relation between center symmetry and other non-perturbative
phenomena occurring in Yang-Mills theories, including the confining
mechanism itself, is an open issue which still needs to be clarified.
A useful theoretical tool, in this respect, is represented by trace
deformation, which was introduced in Refs.~\cite{Unsal:2008ch,
  Myers:2007vc}. For the case of the gauge group $SU(3)$ that will be
studied in this paper, it consists in adding a term proportional to
$|\mathrm{Tr}P(\vec{x})|^2$ to the YM action density.  The rationale
behind this choice is that such a term is invariant under center
symmetry and, if its coefficient is chosen with the appropriate sign,
it disfavors non-vanishing values of the trace of the Polyakov
loop. Such a term compensates analogous ones that appear in the finite
temperature effective potential \cite{Gross:1980br}, and its practical
effect is that of increasing the deconfinement temperature or,
equivalently, of reducing the critical length of the compactified
direction.  This opens the door to the possibility of studying
confinement and other nonperturbative low-energy properties in a
finite temperature setup in which semiclassical methods can provide
solid predictions~\cite{Unsal:2008ch}.

The confinement mechanism at high temperature in the deformed theory,
in what we will call the ``reconfined phase'', could clearly be
different from the one of the original, undeformed YM theory: in the
deformed case confinement is essentially enforced explicitly by a term
in the Lagrangian, rather than emerging as a dynamical property of the
theory. In spite of this, a number of lattice studies showed that the
standard confined phase at $T=0$ in YM theory and the reconfined
high-temperature phase of the deformed YM theory are remarkably
similar, not only from a qualitative point of view but also
quantitatively.  Numerical simulations performed with gauge group
$SU(N)$ both for $N=3$ \cite{Bonati:2018rfg} and $N=4$
\cite{Bonati:2019kmf} showed that the $\theta$-dependence in these two
phases is the same, i.e., the values of the topological susceptibility
and the coefficient of the $\theta^4$ term in the effective action are
the same in standard $T=0$ YM theory and in the reconfined, finite-$T$
phase of the deformed YM theory. Very recently, it was shown that also
the mass of the lowest glueball state is the same in these two
phases~\cite{Athenodorou:2020clr}.

In previous studies the main focus was on the investigation of the
reconfined phase, far from the reconfining phase transition. In this
paper we want to investigate, by lattice simulations, the properties
of the deformed theory in the deconfined phase and around the
transition, studying in particular the possible similarities or
differences with respect to the standard deconfined/confined
transition in YM theory. More precisely, after a finite size scaling
analysis aimed at clarifying the location and the order of the phase
transition, we will concentrate on two aspects that are tightly
connected to the mechanisms of confinement and chiral symmetry
breaking: monopole condensation and Dirac mode localization. As we
will show, some aspects are similar to what happens around the usual
confinement/deconfinement transition, but not all of them; this could
help to highlight some features of confinement and of the deformed YM
theory.

The study of monopoles in lattice YM theories originates from the idea
that color confinement can be a consequence of the condensation of
magnetic degrees of freedom (dual superconductor
scenario)~\cite{'thooft:1975, Mandelstam:1974pi}. Several strategies
have been pursued in order to test this scenario, which go from the
computation of the expectation value of a magnetically charged
operator \cite{DelDebbio:1995yf, DiGiacomo:1997sm, DiGiacomo:1999yas,
  DiGiacomo:1999fb, Carmona:2001ja, Carmona:2002ty, DElia:2005sfk,
  Bonati:2011jv, Cea:2001an} to the extraction of the effective action
of the monopoles \cite{Chernodub:1996ps, Chernodub:2003pu}.  Another
possibility, which is the one we will follow in this work, is to study
the behavior at the transition of thermal
monopoles~\cite{Ejiri:1995gd, Chernodub:2006gu,
  Chernodub:2007cs,DAlessandro:2007lae,
  Liao:2008jg,Ratti:2008jz,DAlessandro:2010jdd,
  Chernodub:2009hc,Bornyakov:2011th, Bornyakov:2011eq, Braguta:2012zz,
  Bonati:2013bga}, i.e., the monopoles whose currents wrap around the
compactified direction \cite{Ejiri:1995gd,Chernodub:2006gu,
  Chernodub:2007cs}.  By studying the space-time configurations of
these monopole currents it is possible to extract an effective
chemical potential $\mu$ for the monopoles \cite{DAlessandro:2010jdd,
  Bonati:2013bga}, whose condensation is signaled by the vanishing of
$\mu$.

Systematic numerical investigations of the spectrum of the Dirac
operator in lattice gauge theories appeared only recently, due to the
high computational complexity of this task (see the reviews
Ref.~\cite{deForcrand:2006my} for results at zero temperature, and
Ref.~\cite{Giordano:2014qna} for the nonzero temperature case). The
existence of a relation between the localization properties of the
Dirac modes and the confinement properties of gauge theories is
however by now a well established fact in a variety of models,
including QCD with different fermionic
discretizations~\cite{GarciaGarcia:2006gr, Kovacs:2009zj,
  Kovacs:2012zq, Cossu:2016scb}, and other QCD-like
theories~\cite{GarciaGarcia:2005vj, Kovacs:2010wx, Giordano:2016nuu,
  Kovacs:2017uiz, Giordano:2019pvc, Vig:2020pgq}.  In the deconfined
phase in the trivial center sector of the Polyakov loop (i.e., the
real sector),\footnote{This is the center sector that would be
  selected by dynamical fermions. In the pure gauge case it is
  selected ``by hand'' when studying the localization properties of
  the Dirac modes.} the
lowest-lying Dirac modes are localized (i.e., their typical space-time
size does not grow with the lattice volume), up to a
temperature-dependent critical point in the spectrum, $\lambda_c$,
known as ``mobility edge''. At the mobility edge, a continuous phase
transition (Anderson transition) takes place in the
spectrum~\cite{Giordano:2013taa,Giordano:2019pvc,Ujfalusi:2015nha},
and Dirac modes become delocalized above $\lambda_c$. As the
temperature is decreased, $\lambda_c$ approaches zero, and eventually
vanishes in the confined phase, where all the low-lying Dirac modes
are extended. In particular, this scenario has been tested numerically
for $SU(3)$ YM using both the staggered Dirac
operator~\cite{Kovacs:2017uiz} and the overlap Dirac
operator~\cite{Vig:2020pgq}. It has been
argued~\cite{Bruckmann:2011cc, Giordano:2015vla, Giordano:2016cjs}
that the presence or absence of localized modes is related to the
different behavior of the Polyakov loop in the deconfined and confined
phases of the theory. Since in the reconfined phase the Polyakov-loop
expectation value vanishes, one would expect only delocalized Dirac
modes. On the other hand, the reasons for its vanishing are different
than in the usual confined phase, and it is worth checking whether
this plays a role or not.

The paper is organized as follows. In Sec.~\ref{sec:numsetup} we
describe the numerical setup and the lattice observables which are
investigated in this study. Sec.~\ref{sec:numresults} contains our
numerical results: we first present a finite size scaling analysis of
the transition between the deconfined and the reconfined phases of the
deformed $SU(3)$ theory; then we study the properties of thermal
monopoles and the localization properties of the low-lying Dirac modes
around the transition.  Finally in Sec.~\ref{sec:concl}, we draw our
conclusions and present some outlooks.

\section{Numerical setup}
\label{sec:numsetup}

The action of trace-deformed $SU(3)$ YM reads
\begin{equation}
S^{\mathrm{def}} = S_{\rm YM} + h \int | \mathrm{Tr}P (\vec{x}) | ^2 \mathrm{d}^3 x\ ,
\label{eq:def_action}
\end{equation}
where $S_{\rm YM}$ is the standard YM action, $\vec{x}$ denotes a
point on a hyperplane perpendicular to the compactified direction
(i.e., the spatial coordinates of a point at a fixed time) and $h$ is
the deformation coupling.  Since each configuration is weighted in the
Euclidean path integral by $e^{-S^{\mathrm{def}}}$, positive values of
the parameter $h$ disfavor nonvanishing values of
$\mathrm{Tr}P(\vec{x})$.

A possible discretization of the action 
in Eq.~\eqref{eq:def_action} is the following,
\begin{equation}
S^{\mathrm{def}} = S_{\rm YM}^{\rm (W)} + h \sum_{\vec{n}} |\mathrm{Tr}P (\vec{n})| ^2\ ,
\label{eq:lattice_action}
\end{equation}
where $S_{\rm YM}^{\rm (W)}$ is the usual Wilson action
\cite{Wilson:1974sk} with bare coupling $\beta$, and $\vec{n}$ is the
lattice analogue of the variable $\vec{x}$ in
Eq.~\eqref{eq:def_action}.  In our numerical simulations we used the
lattice action Eq.~\eqref{eq:lattice_action}, updating the spatial
links with standard heatbath and overrelaxation
algorithms~\cite{Creutz:1980zw, Kennedy:1985nu, Creutz:1987xi},
implemented for $SU(3)$ using the Cabibbo-Marinari
procedure~\cite{Cabibbo:1982zn}; instead, temporal links were updated
with a Metropolis algorithm~\cite{Metropolis:1953am}, since they
appear non-linearly in the deformation term.

\subsection{Thermal monopoles}

The identification of Abelian magnetic monopoles in non-Abelian gauge
theories goes through a procedure known as Abelian
projection~\cite{tHooft:1981bkw}, which consists in a partial gauge
fixing which breaks the gauge symmetry $SU(N)$ down to $U(1)^{N-1}$.
In this work we used a variant of what is usually called Maximal
Abelian Gauge (MAG).  Standard MAG consists in the maximization of the
sum of the squared moduli of the diagonal elements of all link
matrices~\cite{Brandstater:1991sn}: it is not affected by significant
lattice artefacts (see, e.g., the discussion in
Ref.~\cite{Bonati:2013bga}) and satisfies the Dirac quantization
condition~\cite{Bonati:2010tz, Bonati:2010bb}.  However, this
procedure presents some drawbacks: for example, it leaves a residual
permutation symmetry between the different $U(1)$ subgroups, thus
preventing an unambiguous identification of the different monopole
species when using more than two colors. For this reason a variation
of the original MAG was put forward in Ref.~\cite{Stack:2001hm}, and
we will follow the specific implementation proposed in
Ref.~\cite{Bonati:2013bga}, to which we refer for further details
about the gauge fixing algorithm.

Once the gauge is fixed, it is possible to extract the Abelian components
$a^{(i)}_{\mu}$ of each link (with $i=1,\ldots ,N-1$ denoting the monopole species)
and compute the associated DeGrand-Toussaint current $m_{\mu}^{(i)}$ \cite{DeGrand:1980eq}, 
which is defined as
\begin{equation}
m_{\mu}^{(i)} = \frac{1}{2\pi}\epsilon _{\mu\nu\rho\sigma}
\hat{\partial}_{\nu}\bar{\theta}^{(i)}_{\rho\sigma}\ , 
\label{eq:degrand_current}
\end{equation}
where $\hat{\partial}_{\nu}$ is the forward lattice derivative, while
$\bar\theta^{(i)}_{\rho\sigma}$ is derived from the plaquette
$\theta^{(i)}_{\rho\sigma}$ computed using the Abelian fields
$a^{(i)}_{\mu}$ as follows:
\begin{equation}
\theta^{(i)}_{\mu\nu} = \bar{\theta}^{(i)}_{\mu\nu} + 2\pi n_{\mu\nu}^{(i)}\ , \quad 
\bar{\theta}^{(i)}_{\mu\nu}\in [0,2\pi)\ ,\quad n_{\mu\nu}^{(i)}\in\mathbb{Z}\ .
\end{equation}
In the lattice setup $n_{\mu\nu}^{(i)}$ is the analogue of the Dirac string
piercing a plaquette, and this decomposition of $\theta^{(i)}_{\mu\nu}$ is
needed to identify violations of the Abelian Bianchi identity (hence
monopoles): the flux across any closed surface of the magnetic current
constructed using just $\theta^{(i)}_{\mu\nu}$ would indeed identically vanish.

Due to the topological conservation law,
$\partial_{\mu} m^{(i)}_{\mu} = 0$, monopole currents form closed
loops, and we identify thermal monopoles (anti-monopoles) with the
monopole currents which have a non trivial positive (negative) winding
number in the temporal direction. Such currents are interpreted as the
paths of real (instead of virtual) magnetically charged
quasi-particles populating the thermal medium~\cite{Ejiri:1995gd,
  Chernodub:2006gu, Chernodub:2007cs, DAlessandro:2007lae}. Moreover,
in analogy with the path-integral formulation for a system of
identical particles, currents which close after wrapping $k$ times
around the thermal circle are interpreted as a system of $k$ thermal
monopoles undergoing a cyclic permutation~\cite{DAlessandro:2010jdd,
  Bonati:2013bga}: as for a system of identical bosons, the occurrence
of such cycles is the relevant observable which can be used to
investigate the possible condensation of the thermal
particles~\cite{DAlessandro:2010jdd}.

In our study of thermal monopoles we measured the density of
$k$-cycles $\rho_k$, defined as
\begin{equation}
\rho_k \equiv \frac{N_{\mathrm{wrap},k}}{V_s},
\end{equation}
where $V_s=L^3$ is the spatial volume and $N_{\mathrm{wrap},k}$ is the
number of monopole currents wrapping $k$ times around the thermal
circle. From such quantities, the total density of thermal monopoles
can be derived:
\begin{equation}
	\rho = \sum_k k\, \rho_k \, .
\label{eq:total_mon_den}
\end{equation}

\subsection{Dirac operator discretization and spectrum computation}

To investigate the localization properties of fermionic modes we
employed the staggered discretization of the continuum Dirac operator,
which is defined as follows,
\begin{equation}
  \begin{aligned}
&    	a D^{\mathrm{st}}(n,m) \\ &= \frac{1}{2}\sum _{\mu =1}^{4}
\eta _{\mu}(n) (U_{\mu}(n)\delta _{n+\hat{\mu}, m}  
- U_{\mu}(n-\hat{\mu})^\dag\delta _{n-\hat{\mu}, m}),
  \end{aligned}
\label{eq:stag_operator}
\end{equation}
where $\eta_{\mu}(n)$ are the staggered phases, $U_{\mu}(n)$ is the
link variable at site $n$ in direction $\mu$, and $a$ is the lattice
spacing. Since $D^{\mathrm{st}}$ is anti-Hermitian, its eigenvalues
are purely imaginary. To fix the notation, we then write the
eigenvalue equation as $a D^{\mathrm{st}}\psi_n= i a\lambda_n \psi_n$,
$\lambda_n\in\mathbb{R}$, where $n$ labels the discrete modes of
$D^{\mathrm{st}}$ in a finite volume. Since $D^{\mathrm{st}}$
anticommutes with $(-1)^{\sum_\mu n_\mu}$, its spectrum is symmetric
with respect to $\lambda=0$, so that it suffices to consider
$\lambda_n\ge 0$.

For each choice of the bare parameters we have computed the lowest
part of the spectrum and the corresponding eigenvectors on a set of
uncorrelated configurations, making use of the ARPACK
routine~\cite{lehoucq1998arpack}. Diagonalization was carried out
after two steps of stout smearing on the gauge configuration, with
stout parameter $\rho = 0.15$ (see Ref.~\cite{Morningstar:2003gk} for
details).

\section{Numerical Results}
\label{sec:numresults}

We have performed simulations for a fixed value of $N_t = 6$ and $\beta = 6.0$,
changing only the deformation coupling $h$, or the spatial extent for finite
size scaling purposes: this is sufficient for a first assessment of the
properties of the reconfining transition, even if no continuum extrapolation is
possible.  Using the interpolation formula for the Sommer parameter $r_0$ of
Ref.~\cite{Necco:2001xg} and the phenomenological value $r_0=0.5\,\mathrm{fm}$, we
infer that our setup corresponds to a temperature $T = 1 / (N_t a(\beta))
\simeq 360$~MeV, meaning that at $h = 0$ the system is in the deconfined phase:
indeed, for $N_t=6$ the critical coupling for deconfinement is $\beta _c =
5.8941(5)$~\cite{Fingberg:1992ju}.

\subsection{The reconfinement transition}

In this section we study the reconfinement transition, by identifying the
critical value $h_c$ such that for $h>h_c$ (at $\beta=6.0$ on $N_t=6$ lattices)
the system is in the reconfined phase, and by investigating the order of the
reconfinement transition.
 
To identify $h_c$ we monitor the behavior of the susceptibility of the
modulus of the Polyakov loop,
\begin{equation}
\chi_P = V_s \left( \langle |\mathrm{Tr}P|^2 \rangle - \langle |\mathrm{Tr}P|\rangle ^2 \right)\ ,
\label{eq:susceptibility}
\end{equation}
where $V_s=L^3$ is the spatial volume of the lattice and we used the shorthand
\begin{equation}
\mathrm{Tr} P \equiv \frac{1}{V_s}\sum_{\vec{n}} \mathrm{Tr}P(\vec{n})\ .
\end{equation}

\begin{figure}[t]
\includegraphics[width=0.9\columnwidth, clip]{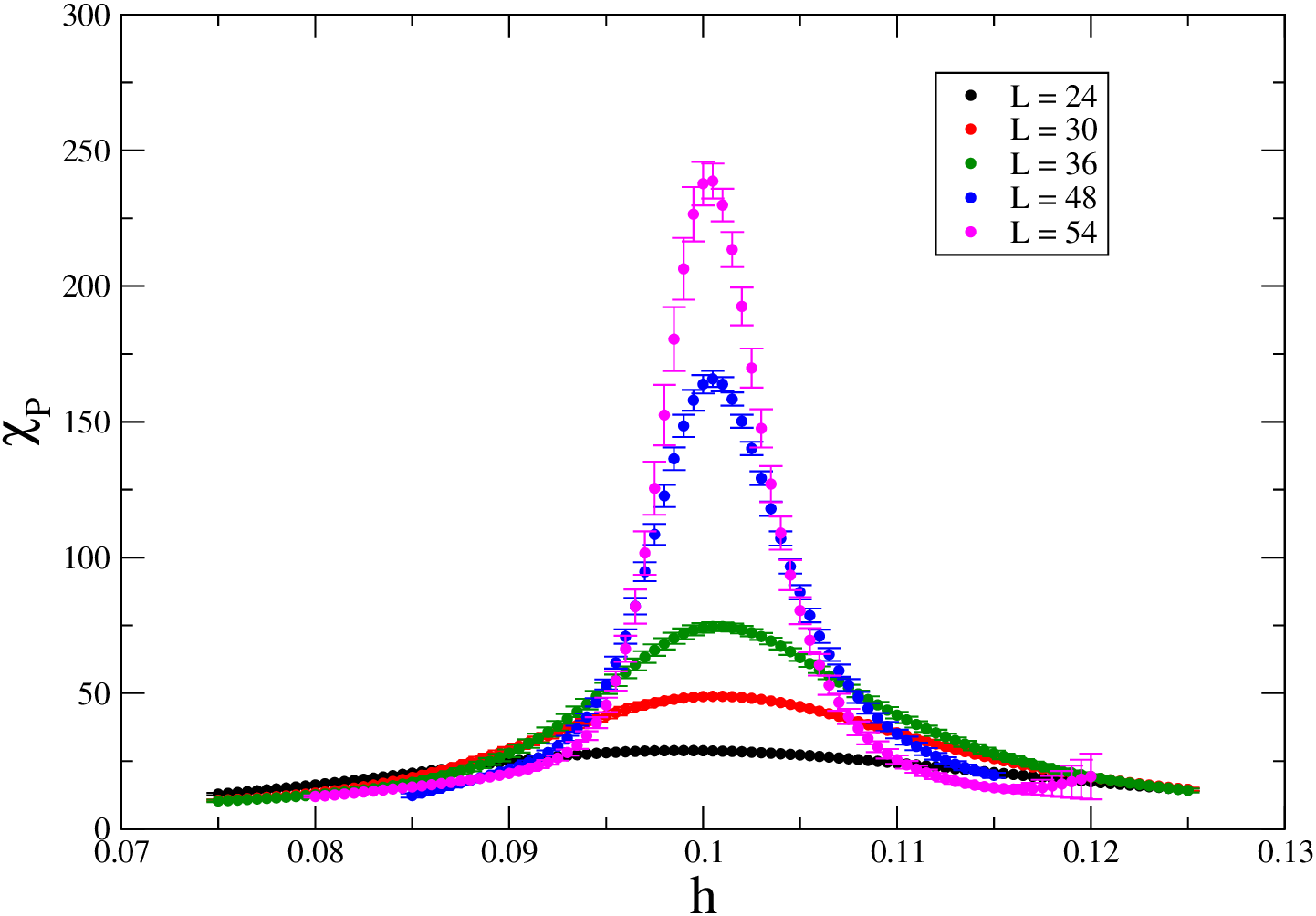}
\caption{Polyakov loop susceptibility $\chi_P$ as a function of $h$ for several
spatial volumes, $\beta = 6.0$ and $N_t=6$.} 
\label{fig:all_volumes_susc}
\end{figure}

For each spatial size $L$, data for $\chi_P$ have been produced using several
values of $h$ and analyzed using the multiple histogram
method~\cite{Ferrenberg:1989ui}.  The final results of this analysis are
reported in Fig.~\ref{fig:all_volumes_susc}: the susceptibility develops a peak
that gets higher and narrower when increasing $L$, signaling the presence of a
phase transition. Since the breaking pattern at reconfinement is the same
$\mathbb{Z}_3 \ \rightarrow \ \mathrm{Id}$ of the standard deconfinement phase
transition, the universality argument of Refs.~\cite{Svetitsky:1982gs,
Yaffe:1982qf} predicts a first order phase transition also in this case.

Finite size scaling (FSS) for first order phase transitions in a
translation invariant setup predicts the following scaling behavior
for the susceptibility of the order parameter,
\begin{equation}\label{eq:chi_scaling_th}
\chi_P(h)=L^{\gamma/\nu}f\Big(L^{1/\nu}(h-h_c)\Big)\ ,
\end{equation}
with effective critical exponents $\gamma=1$ and $\nu=1/3$
\cite{Fisher:1982xt, CLB-86, VRSB-93, LK-91}.  We thus first of all
verified that the height of the peaks of the susceptibility scales as
expected with the volume, fitting the maxima with the functional form
\begin{equation}
\chi_P ^{\mathrm{max}} = a L^{3b}\ ,
\label{eq:fit_maximum_susc}
\end{equation}
where $a$ and $b$ are fit parameters. The best fit value $b=0.98(3)$ is indeed
consistent with the theoretical expectation $b=1$. 

\begin{figure}[t]
\includegraphics[width=0.9\columnwidth, clip]{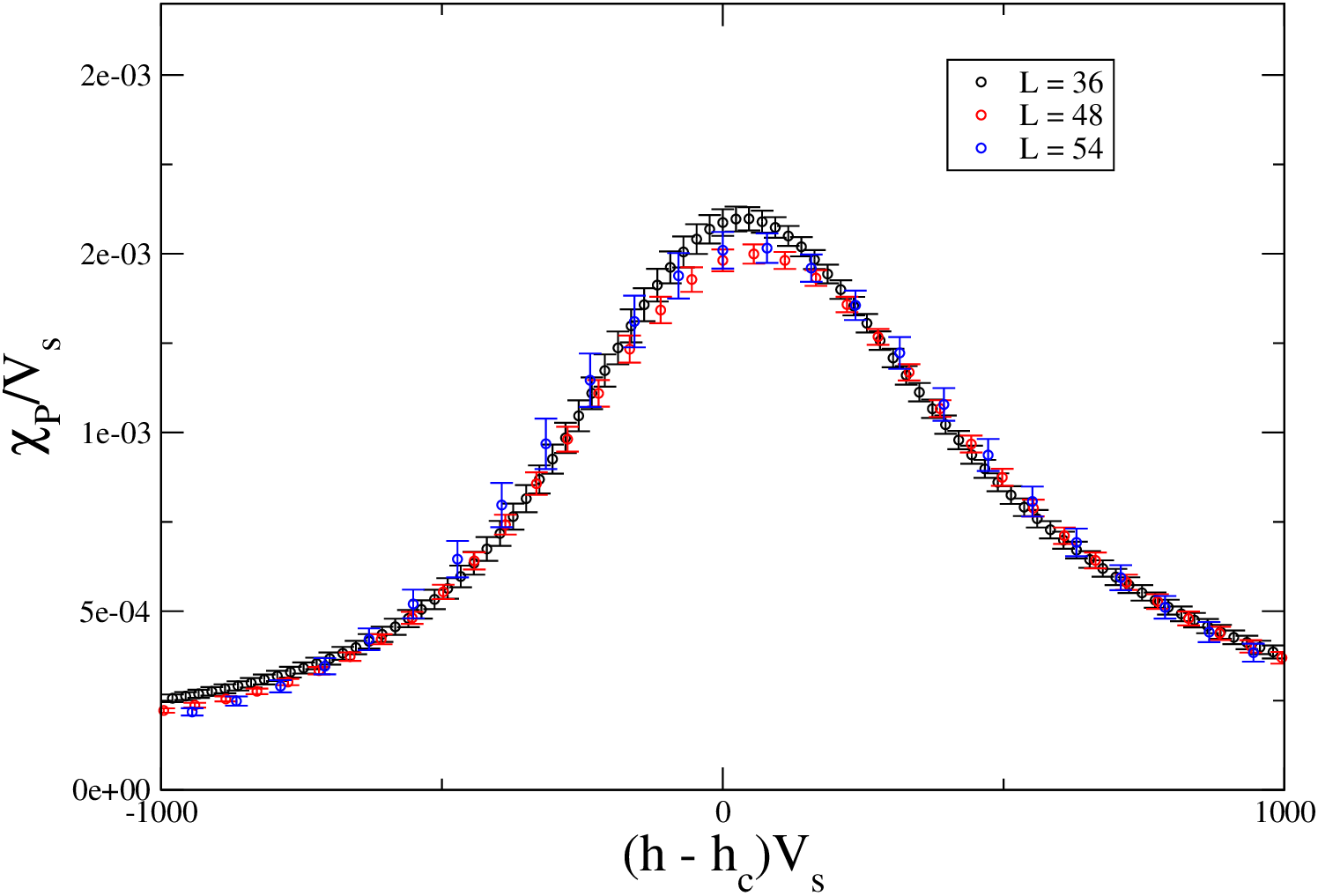}
\caption{Finite size scaling of the Polyakov loop
susceptibility $\chi_P$ according to Eq.~\eqref{eq:chi_scaling_th},
after fixing $\nu = 1/3$, $\gamma = 1$ and $h_c=0.1$.}
\label{fig:fss}
\end{figure}

We then verified that the data are well described by the scaling ansatz in
Eq.~\eqref{eq:chi_scaling_th}, using the critical exponents appropriate for a
first order phase transition: the corresponding collapse plot is shown in
Fig.~\ref{fig:fss} and fully supports the transition being first order. By
varying $h_c$ in this scaling plot we estimated $h_c=0.100(2)$: for $h_c$
outside this interval the peaks of the data corresponding to $L=36$ and $L=54$
data sets become clearly separated from each other.

\subsection{Thermal monopole condensation} 
 
\begin{figure}[t!]
\includegraphics[width=0.9\columnwidth, clip]{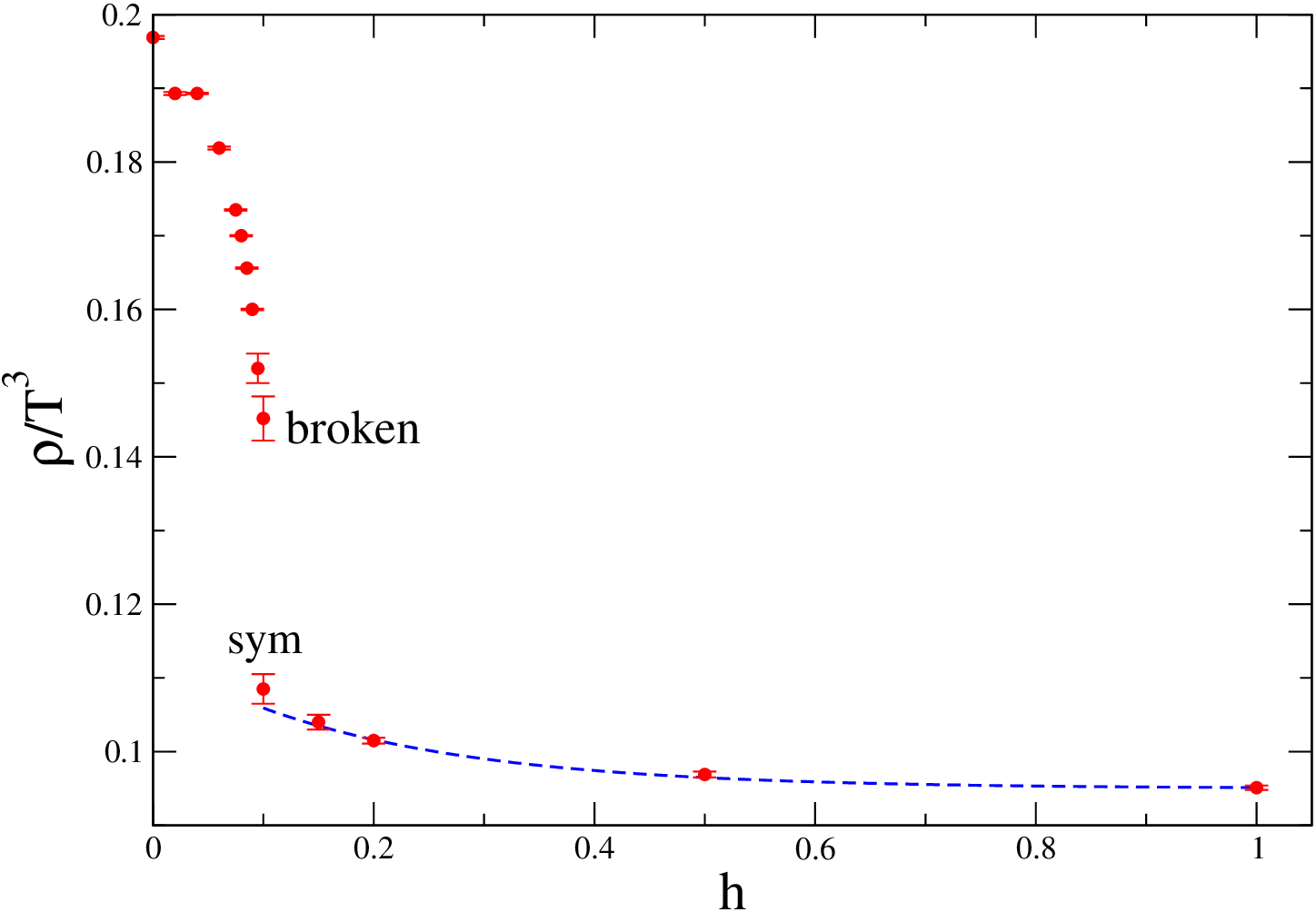}
\caption{$\rho /T^3$ computed using different values of the deformation
parameter $h$ on a $54^3\times6$ lattice at $\beta = 6.0$. 
The two determinations reported for the transition point 
$h_c=0.1$ have been obtained by 
dividing the sample of configurations into two subsamples according
to the realization of center symmetry (broken/unbroken).}
\label{fig:rho_over_t3}
\end{figure}

Let us discuss first the results obtained for the total density of thermal
monopoles, defined in Eq.~(\ref{eq:total_mon_den}).  We consider, in
particular, the dimensionless quantity $\rho /T^3$, which is reported as a
function of $h$ in Fig.~\ref{fig:rho_over_t3}.  The behavior of $\rho/T^3$
presents a significant difference with respect to what happens when approaching
the usual confinement/deconfinement transition~\cite{DAlessandro:2010jdd,
Bonati:2013bga}: in that case, $\rho / T^3$ grows as one moves from the
deconfined towards the confined phase, and decreases at high $T$, where it
approximately follows the behavior predicted by the perturbative analysis
\cite{Liao:2006ry, Giovannangeli:2001bh}, $\rho /T^3 \approx
1/(\mathrm{log}(T/\Lambda _{\mathrm{eff}}))^3$, where $\Lambda _{\mathrm{eff}}$
is some effective energy scale. 

On the contrary, what we observe from Fig.~\ref{fig:rho_over_t3} is
that $\rho/T^3$ steeply decreases approaching the reconfined phase,
has a big negative jump in correspondence of the first order phase
transition, and then continues a slow decrease also in the reconfined
phase. Results in the reconfined phase suggest the approach to a
constant value in the large-$h$ limit; indeed, a best fit of results
in the reconfined phase to a function
\begin{equation}
\frac{\rho}{T^3} = \frac{\rho_{h = \infty}}{T^3} + A\, e^{-h/\bar h}
\end{equation}
returns $\rho_{h=\infty}/T^3 = 0.095(1)$ and $\bar h = 0.20(5)$, 
with $\chi^2/\textrm{d.o.f.} = 2.8/2$.

The fact that $\rho/T^3$ decreases approaching reconfinement may seem
at odds with the possibility that monopoles condense at the
transition. However, this is emblematic of a possible
misinterpretation of the meaning of condensation, which is not
necessarily related to an increase in the overall density, but rather
to the appearence of a non-zero density of particles in the
zero-momentum state. For a boson gas, such condensation is signalled
in the path-integral formulation by a critical behavior of the density
of $k$-cycles $\rho_k$ as a function of $k$, and this is indeed a
possible criterion proposed to study thermal monopole
condensation~\cite{cristoforetti,DAlessandro:2010jdd}.  Therefore we
now turn to this kind of analysis, which will finally show that, in
fact, monopoles condense at $h_c$, in spite of the decrease of
$\rho/T^3$.

In the path integral representation of the partition function of bosonic
particles, quantum effects can be associated with particle paths undergoing a
permutation around the thermal circle, hence to trajectories wrapping multiple
times around the temporal direction, which represent the cycle decomposition of
the corresponding permutation~\cite{fey1,fey2,elser,ceperley}.  In particular,
the densities $\rho_k$ are expected to behave as follows,
\begin{equation}
	\rho _k \propto \frac{e^{-\hat{\mu}k}}{k^{\alpha}}\ ,
\label{eq:density_bec}
\end{equation}
where $\hat{\mu} \equiv - \mu /T$, with $\mu$ the chemical potential,
and $\alpha$ is a coefficient depending on the details of the system,
for instance $\alpha = 5/2$ for non-interacting non-relativistic
bosons. At high temperatures $\hat \mu$ is large and paths with
multiple wrappings are rare, since one approaches Boltzmann
statistics. On the contrary, $\hat \mu$ decreases at low temperatures
and its vanishing signals the occurrence of a critical phenomenon like
Bose-Einstein Condensation (BEC).  Once BEC is reached,
$\hat{\mu} = 0$ and $\rho_k$ follows a power-law behavior.

\begin{figure}[t!]
\includegraphics[width=0.9\columnwidth, clip]{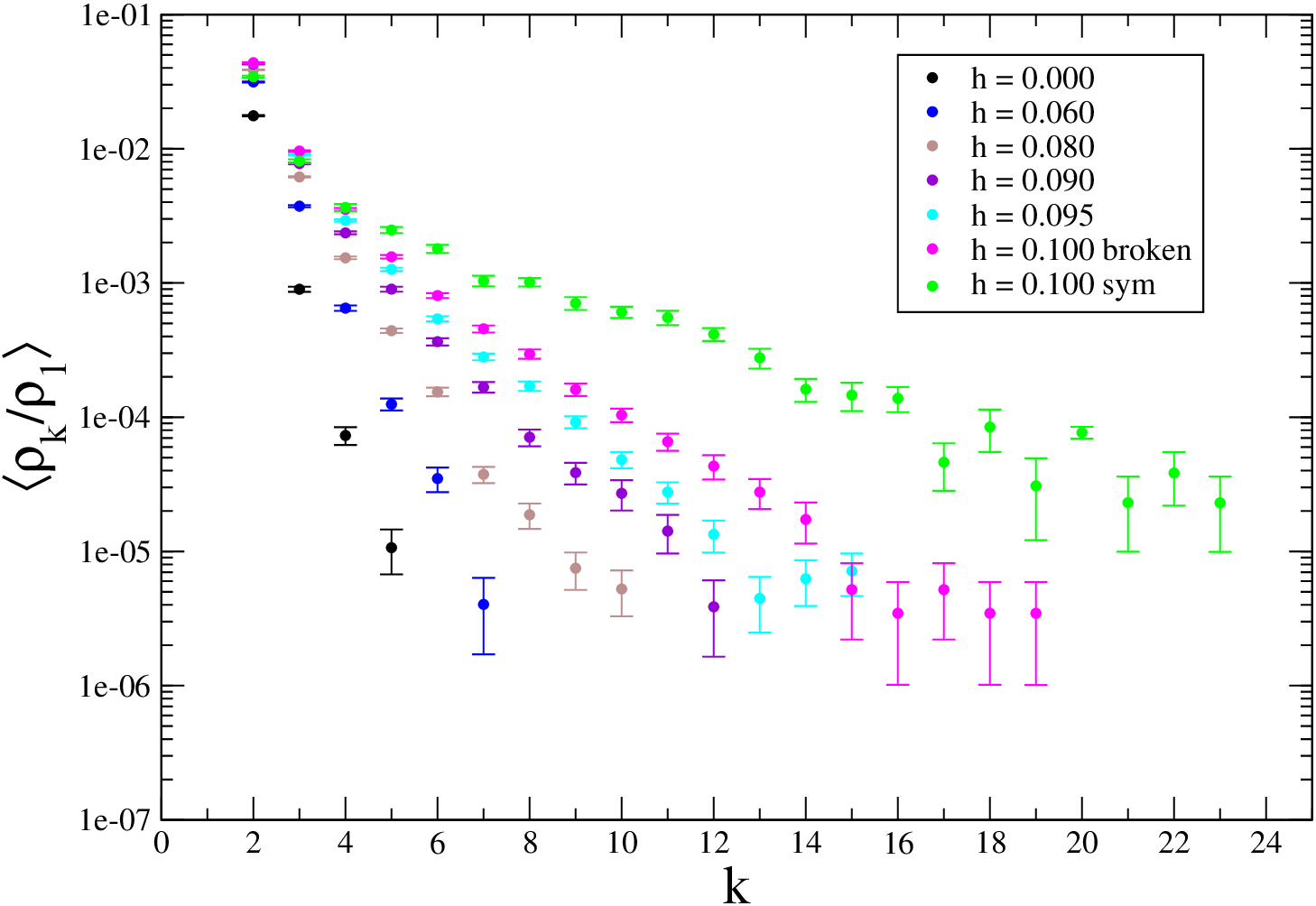}
\caption{$\langle \rho _k / \rho _1 \rangle$ approaching the reconfined phase.
The lattice used is a $54^3\times 6$ at $\beta = 6.0$.}
\label{fig:rhok_over_rho1}
\end{figure}

Results obtained for the ratio $\rho_k /\rho_1$ for different values
of $h$ are reported in Fig.~\ref{fig:rhok_over_rho1}: it is already
quite clear from this figure that the density of monopole trajectories
having multiple wrappings increases as the coupling $h$ grows, i.e.,
approaching the reconfined phase.

\begin{figure}[t!]
\includegraphics[width=0.9\columnwidth, clip]{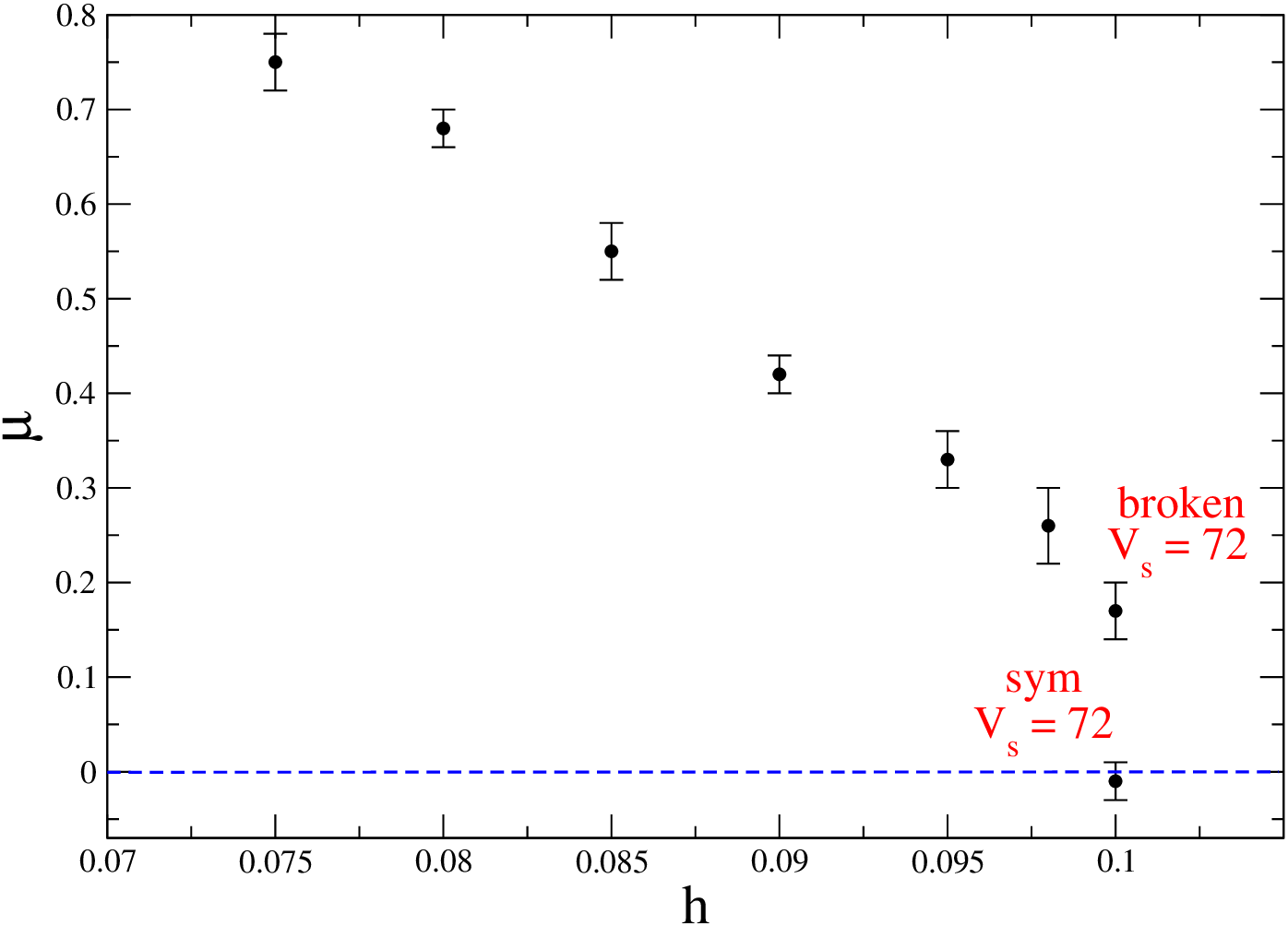}
\caption{The dimensionless chemical potential $\hat{\mu}$ extracted
  at different values of the deformation coupling $h$. The lattice
  used is a $54^3\times 6$ at $\beta = 6.0$.}
\label{fig:mu}
\end{figure}

Next we performed a best fit of $\rho_k/\rho_1$ to
Eq.~(\ref{eq:density_bec}), in order to extract the chemical potential
$\hat{\mu}$ as a function of $h$.  As in
Refs.~\cite{DAlessandro:2010jdd,Bonati:2013bga}, statistics are not
precise enough to obtain an independent determination of $\alpha$,
since different choices, including $\alpha = 5/2$, lead to acceptable
best fits; however, the analysis of
Refs.~\cite{DAlessandro:2010jdd,Bonati:2013bga} shows that this does
not affect the determination of a possible point where $\hat \mu$
vanishes.  Results obtained for $\hat \mu$ fixing $\alpha = 5/2$ are
shown in Fig.~\ref{fig:mu}. The dimensionless chemical potential
decreases as the system moves towards the reconfined phase, however it
seems not to reach the value $\hat{\mu} = 0$ at the critical
$h_c$. This can still be compatible with a monopole condensation
scenario if the chemical potential has a jump, rather than vanishing
continuosly, at $h_c$. This would be consistent with the presence of a
first order transition, which would be stronger, at least for what
concerns monopole condensation, than in the standard case of the
thermal phase transition, where instead $\hat\mu$ does not show any
appreciable jump at the transition~\cite{Bonati:2013bga}. In order to
check this hypothesis, however, one should verify that the behavior of
$\rho_k$ right in the reconfined phase is compatible with a power law
behavior, i.e., with a vanishing value of $\hat \mu$.

In order to clarify this point we performed simulations at $h=h_c$,
dividing the configurations into two different subsamples: the
``symmetric'' configurations where center symmetry is unbroken (i.e.,
$\mathrm{Tr}P\simeq 0$), and the ``broken'' configurations where
center symmetry is broken (i.e., $\mathrm{Tr} P\neq 0$); in order to
make the division sharper and better defined, these additional
simulations were performed adopting a larger spatial size, in
particular a $72^3 \times 6$ lattice. The increase in the relative
occurence of trajectories with a high number of wrappings in the
symmetric configurations, clearly visible in
Fig.~\ref{fig:rhok_over_rho1}, is suggestive of a sudden change in the
behavior of $\rho_k$ as $h$ crosses the critical value. We have then
fitted the two sets of configurations using Eq.~\eqref{eq:density_bec}
with $\alpha = 5/2$.  While a nonzero value of $\hat{\mu}$ is returned
by the fit for the broken configurations, a value compatible with zero
is obtained for the symmetric ones. This supports the monopole
condensation scenario in the reconfined phase. It should however be
noted that since the present quality of the data does not allow to
determine the power-law exponent independently, other possibilities
are not entirely excluded.

\subsection{Localization properties of Dirac modes} 
In this section we study the localization properties of the eigenmodes
of the Dirac operator on the lattice.  As it is known from
Refs.~\cite{Kovacs:2017uiz, Vig:2020pgq}, in the deconfined phase of
undeformed $SU(3)$ YM (i.e., at $h=0$) in the trivial center sector of
the Polyakov loop, the lowest Dirac modes are localized in a finite
spatial region of the lattice, for eigenvalues below a critical
mobility edge, $\lambda _c$, in the spectrum.  Higher modes, above the
mobility edge, are instead delocalized on the whole space. Approaching
the confined phase from higher temperatures, the spectral range where
modes are localized shrinks, i.e., $\lambda_c$ decreases, eventually
vanishing at a temperature compatible with the deconfinement
temperature. Here we want instead to investigate what happens when the
system approaches the reconfined phase in the trace-deformed theory,
starting from the deconfined phase: using the same setup described
above, we have studied the spectrum of the staggered Dirac operator in
a range of deformation couplings $h$ starting from $h<h_c$ and
reaching the transition region. Configurations were restricted to the
trivial center sector, by multiplying the temporal links on the last
time-slice by the appropriate center element if necessary.

The simplest observable sensitive to the localization properties of
the Dirac modes is the so-called Participation Ratio (PR), which
essentially measures the fraction of lattice volume occupied by a
mode. For our purposes it is convenient to use a gauge-invariant
definition of the PR. We have then set
\begin{equation}
  \mathrm{PR}_n = \frac{1}{V_s N_t}
  \left[ \textstyle\sum _{\vec{n},t} \left| \psi
      ^{\dagger}_n(\vec{n},t)\psi_n(\vec{n},t) \right| ^2
  \right]^{-1}, 
\label{eq:pr}
\end{equation}
where $\psi^\dag\psi$ denotes the scalar product in color space.  If
modes in a given spectral range are localized, their PR (averaged over
configurations) will tend to 0 as $1/V_s$ as the volume is increased;
if instead they are delocalized, their PR will tend to a
constant. Equivalently, one can look at the {average spatial ``size''
  of the mode,} $\mathrm{VPR}\equiv V_s\cdot \mathrm{PR}$, which will
tend to a constant or diverge like $V_s$ for localized or delocalized
modes, respectively.

\begin{figure}[t]
\includegraphics[width=0.9\columnwidth, clip]{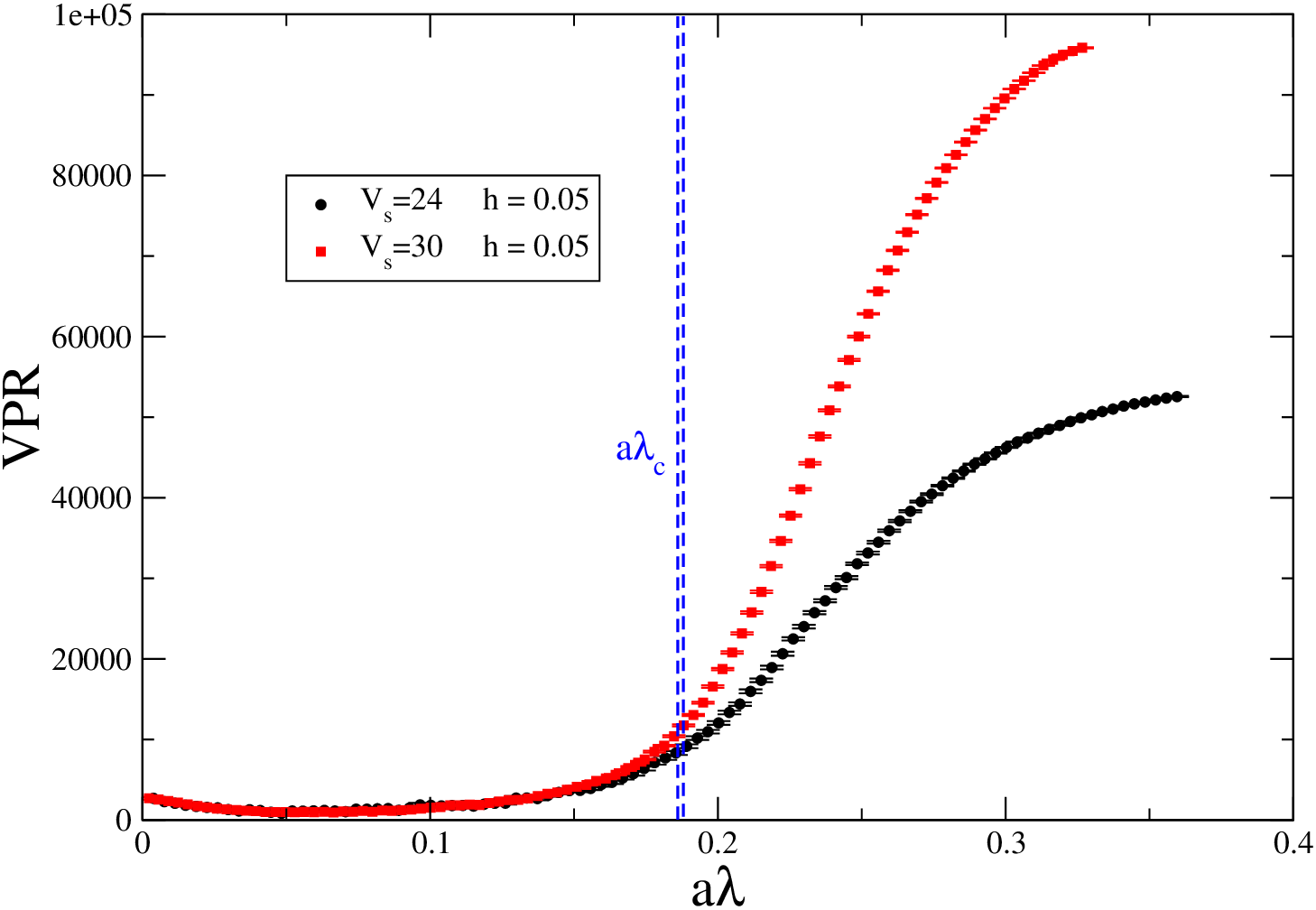}
\includegraphics[width=0.9\columnwidth, clip]{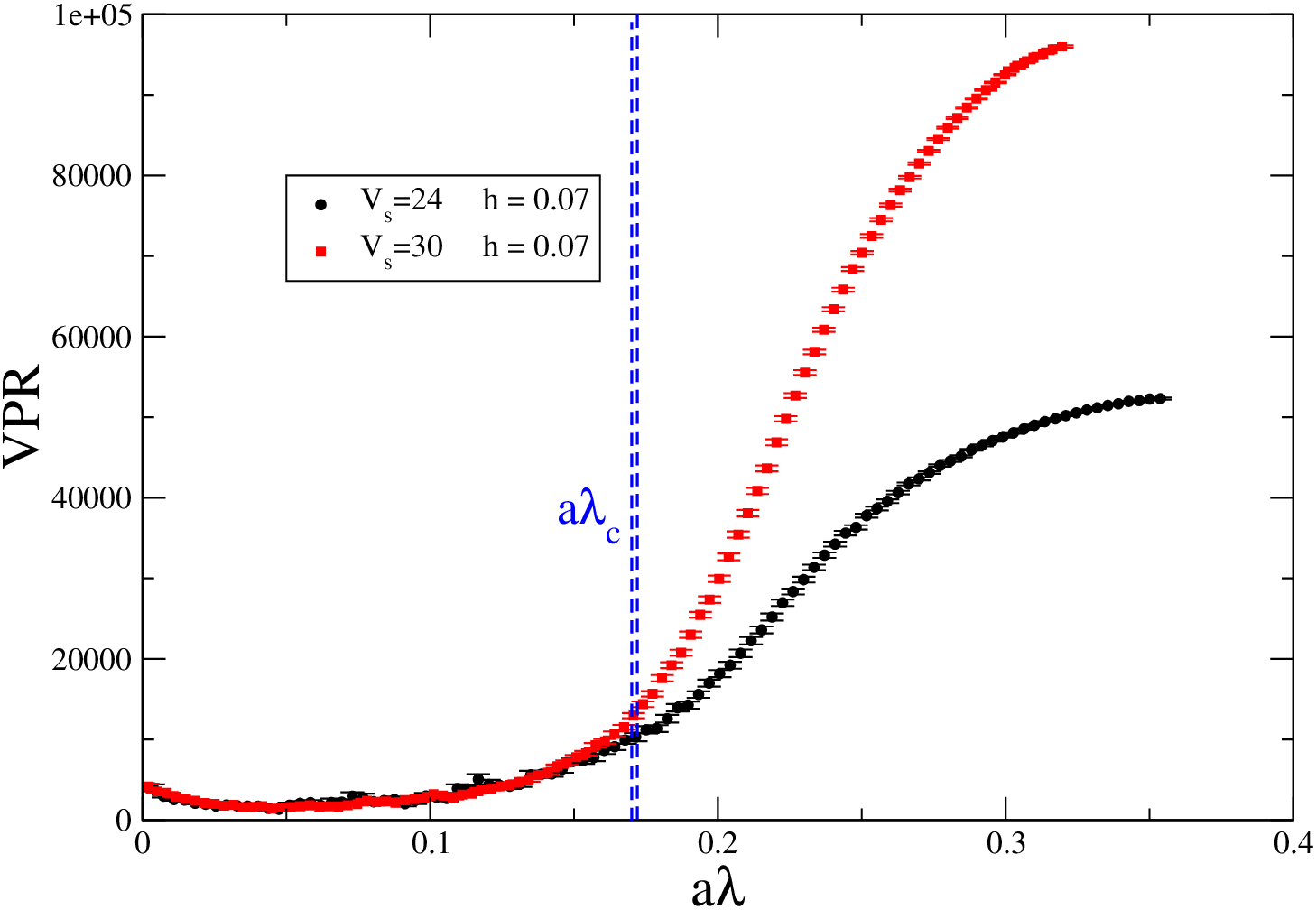}
\includegraphics[width=0.9\columnwidth, clip]{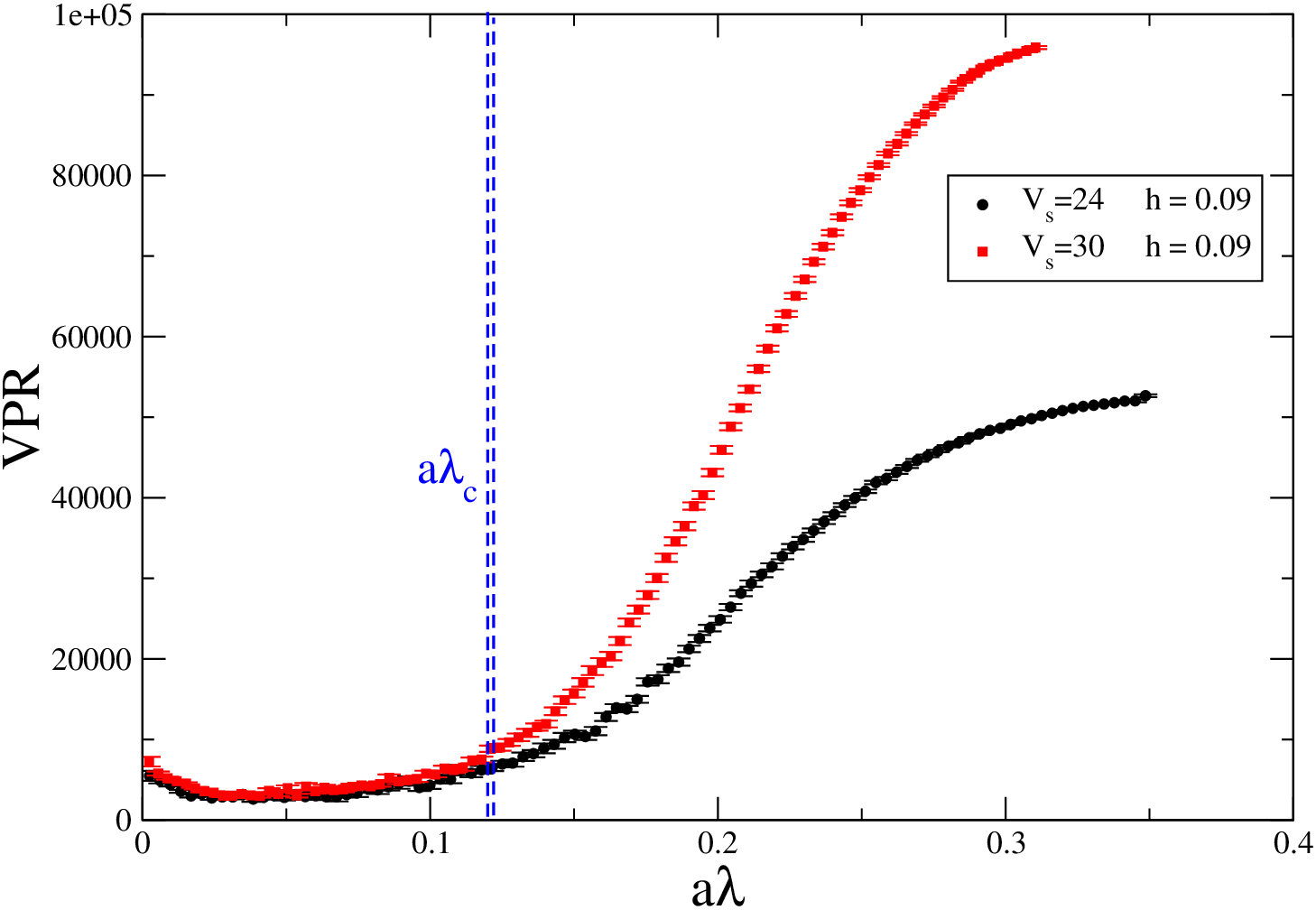}
\caption{Average VPR of the lowest staggered
  eigenmodes for different values of the deformation coupling $h$ below the
  critical one, and for two spatial volumes.} 
\label{fig:vpr_nocrit}
\end{figure}

In Fig.~\ref{fig:vpr_nocrit} we show the VPR of the low Dirac modes
for different lattice setups. We have considered three values of the
deformation coupling, namely $h=0.05, \ 0.07, \ 0.09$, which are less
than the critical coupling $h_c$, and two spatial volumes
$V_s=24^3,30^3$. After dividing the spectrum in small bins, we
averaged the VPR over configurations separately within each bin.  In
Fig.~\ref{fig:vpr_nocrit} one can clearly distinguish two regions. For
the lowest modes the VPR does not change with $V_s$, indicating that
they are localized. Higher up in the spectrum, instead, the VPR of the
bulk modes grows with $V_s$. Assuming the scaling law
$\langle\mathrm{VPR}\rangle_\lambda \sim C(\lambda)
L^{\alpha(\lambda)}$, where $\langle\mathrm{VPR}\rangle_\lambda$ is
the VPR averaged locally in the spectrum, and $\alpha$ is the fractal
dimension of the corresponding modes, one can obtain $\alpha$ by
comparing the two available volumes. For the bulk modes
$\alpha\sim 3$, as expected for delocalized modes. This scenario is
exactly the same found in non-deformed, deconfined
YM~\cite{Kovacs:2017uiz}.

\begin{figure}[t]
\includegraphics[width=0.9\columnwidth, clip]{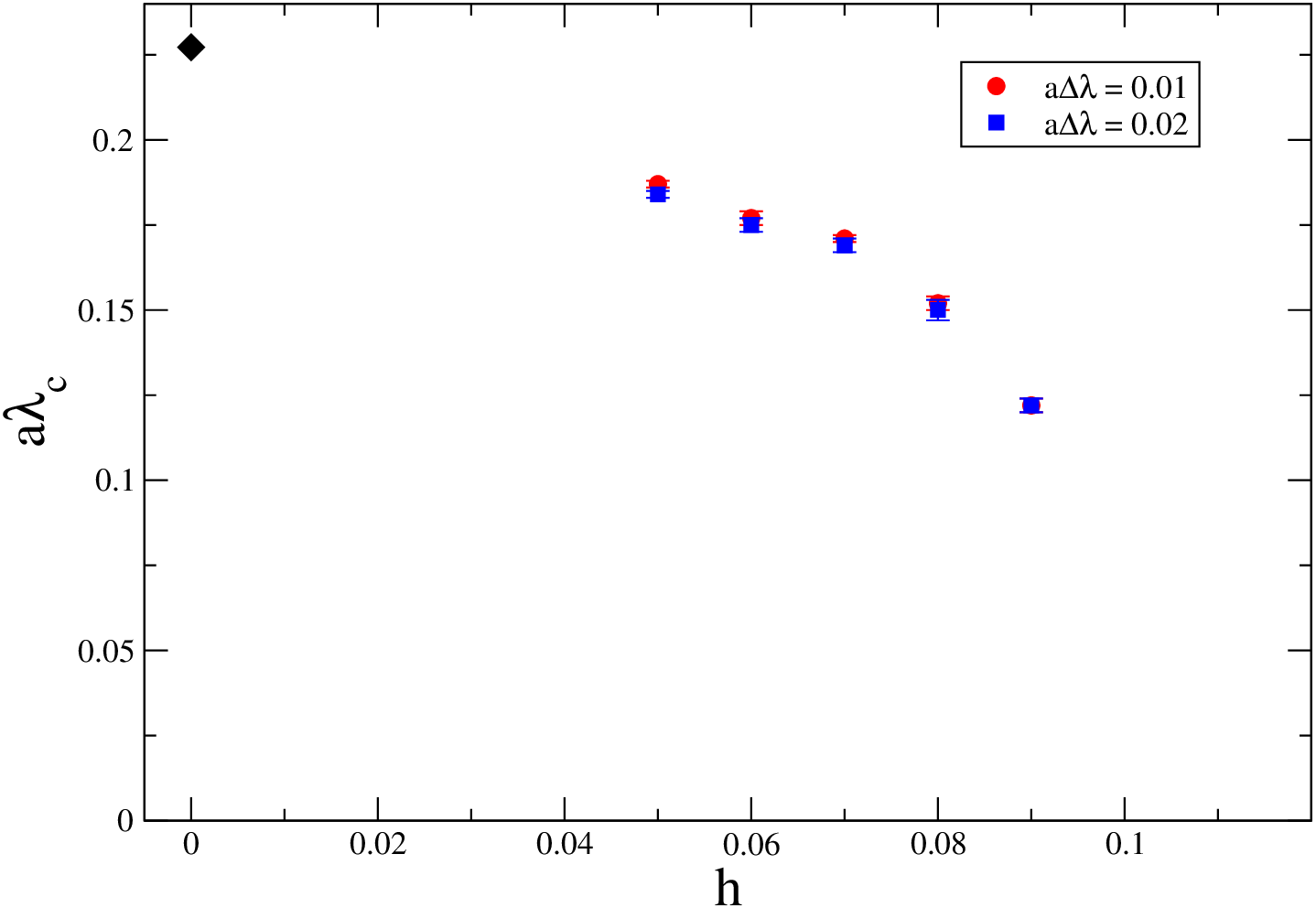}
\caption{Mobility edge, $\lambda_c$, as a function of the deformation
  parameter $h$. The two sets of points correspond to using a
  different bin size $\Delta\lambda$ in the extraction of
  $\alpha(\lambda)$ from the VPR. The black square is $\lambda_c(h=0)$
  obtained by the authors of Ref.~\cite{Kovacs:2017uiz}, and is shown
  for comparison (we thank T.~G.~Kov{\'a}cs and R.~Vig for making
  their results available to us).}
\label{fig:mobedge}
\end{figure}

It is clear from Fig.~\ref{fig:vpr_nocrit} that the region where modes
are localized shrinks as $h$ approaches the critical value $h_c$.  In
order to make this statement more quantitative, we have identified the
mobility edge $\lambda_c(h)$ by comparing the fractal dimension of the
modes with the critical fractal dimension
$\alpha_c = 1.173_{-0.026}^{+0.032}$ found at the mobility edge in the
unitary Anderson model~\cite{Ujfalusi:2015mxa}. The critical behavior
at the mobility edge is in fact expected to be universal and
determined only by the symmetry class of the system in the
classification of Random Matrix Theory, which for the staggered Dirac
operator is the unitary one~\cite{Verbaarschot:2000dy}. In
Fig.~\ref{fig:mobedge} we show $\lambda_c(h)$ as obtained from
$\lambda_c(h)=\alpha_c$ for the various choices of $h$. The tendency
of $\lambda_c(h)$ to decrease as $h\to h_c$ is evident.

\begin{figure}[t]
\includegraphics[width=0.9\columnwidth, clip]{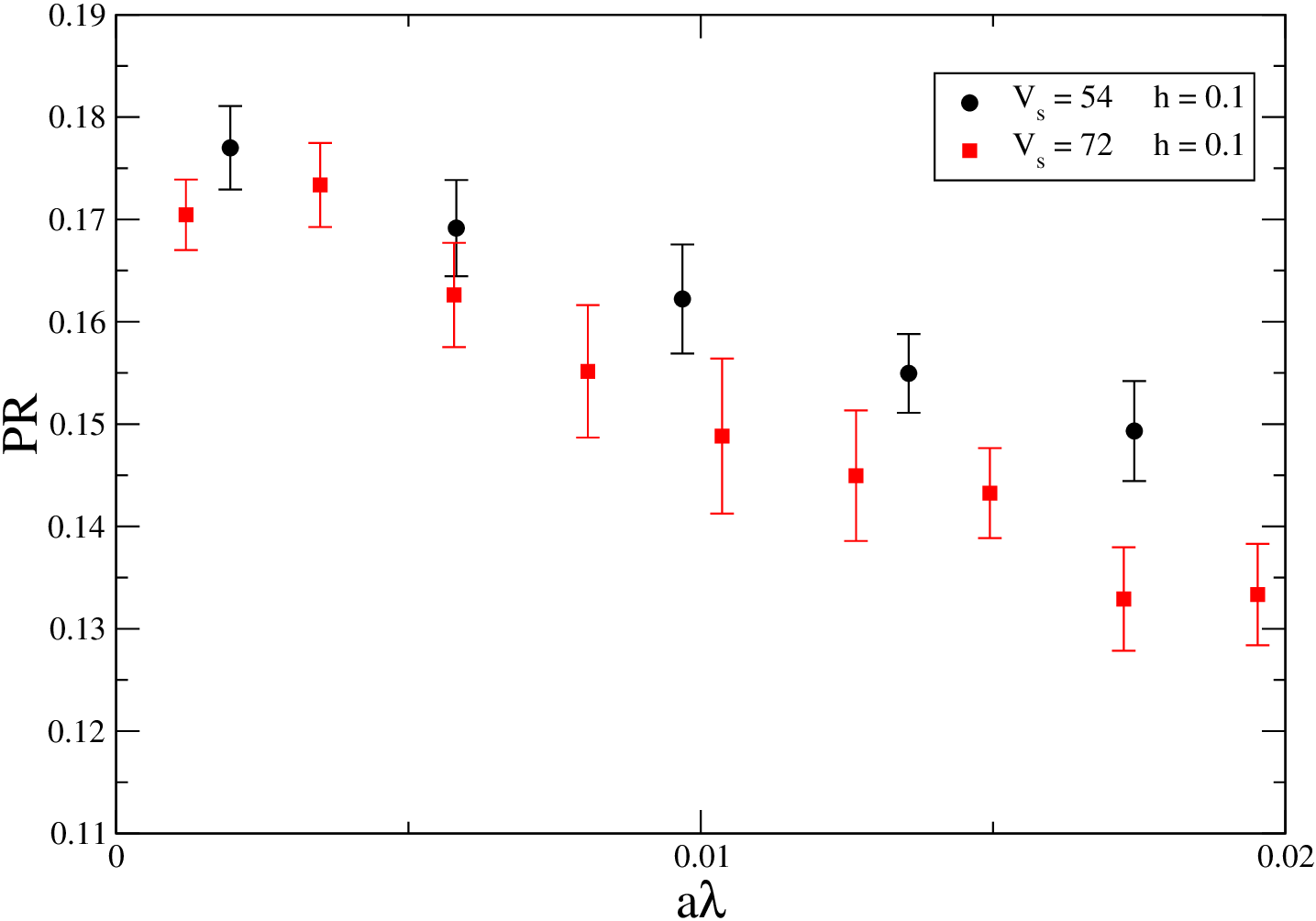}
\caption{Average PR of the lowest staggered
  eigenmodes on ``symmetric'' configurations  at $h=h_c=0.1$.} 
\label{fig:vpr_critical}
\end{figure}

A look at the lowest modes of the staggered operator in the reconfined
phase shows no evidence of localization. More precisely, we looked at
the low modes on a few configurations at $h=0.11$ for $V_s=24^3,30^3$,
and found in the lowest bin ($a\lambda\le 0.01$)
${\rm PR} \sim \, 0.24$ for $L=24$ and ${\rm PR} \sim \, 0.21$ for
$L=30$.  To investigate this issue further, we have generated
configurations at the critical coupling $h_c=0.1$ and divided them in
two subsamples, according to the fate of center symmetry, as we have
done for the analysis of thermal monopoles. For both sets we have
computed the PR on two different spatial volumes, namely $V_s = 54^3$
and $V_s = 72^3$. In Fig.~\ref{fig:vpr_critical} we show our results
for the ``symmetric'' configurations, corresponding to the reconfined
phase. The independence of the PR of the system size confirms the
absence of localized modes in the reconfined phase, all the way down
to $h_c$. Absence of localized modes in the reconfined phase can be
expressed by setting $\lambda_c(h>h_c)=0$.  It is not clear from
Fig.~\ref{fig:mobedge} whether $\lambda_c$ approaches zero
continuously or not as $h\to h_c$: in analogy with what is observed
for monopole condensation, it is reasonable to suppose that there is a
jump also in this case. We would then expect to find localization, and
so a nonzero mobility edge, in the ``broken'' configurations at
$h=h_c$. Unfortunately, the analysis of these configurations still
suffers from sizeable finite-size effects for the available volumes,
and we cannot make a conclusive statement.

\section{Conclusions} 
\label{sec:concl}

In this paper we have studied the reconfinement phase transition of
trace deformed $SU(3)$ YM theory by means of lattice simulations.  In
our analysis we have considered three aspects of the phase transition:
i.)~the order of the transition; ii.)~the behavior of thermal
monopoles; and iii.)~the localization properties of Dirac eigenmodes.
All results were obtained on a lattice with temporal size $N_t=6$ at
$\beta=6.0$, corresponding to an extension in the compactified
direction of approximately $(360 \ \mathrm{MeV})^{-1}$
($0.55~{\rm fm}$).  The order of the transition was determined with a
FSS study, that showed that the reconfinement phase transition is
first-order, exactly as the standard YM thermal deconfinement
transition. This result is not surprising, because both in undeformed
and deformed YM the symmetry breaking pattern is the same, i.e.,
$\mathbb{Z}_3 \ \rightarrow \ \mathrm{Id}$. What the deformation does
is just suppressing local fluctuations of the Polyakov loop (see
Ref.~\cite{Bonati:2018rfg}), forcing the order parameter
$\langle \mathrm{Tr}P \rangle$ to be equal to zero. It would be
interesting, in future studies, to extend this FSS analysis to the
case of a smaller compactification length, in order to see how the
critical value of the deformation coupling changes. Moreover, a
detailed study of the order of the possible phase transition in trace
deformed $SU(N)$ with $N> 3$ would also be very interesting.

Some particular features emerged from the study of thermal monopoles.
We have showed that their normalized density, $\rho/T^3$, decreases
approaching the reconfinement transition and also beyond, reaching a
plateau value in the large-$h$ limit. The fact that the trace
deformation induces a decrease in the monopole density is not
completely unexpected. Abelian magnetic monopoles are localized around
points where two eigenvalues of the corresponding Higgs field
vanish~\cite{tHooft:1981bkw}: had we studied monopoles in the Polyakov
gauge, a decrease in the monopole density would have been a
consequence of the fact that the trace deformation induces a repulsion
among the Polyakov loop eigenvalues.  However, it is natural to expect
that this might have an indirect effect also on monopoles defined in
other Abelian projections, like MAG.  In spite of the decreased
density, thermal monopoles show a behavior compatible with a BEC-like
condensation at reconfinement, as it happens at the standard
confinement/deconfinement phase transition.  It is interesting that
there seems to be a significant jump in observables related to thermal
monopoles at the reconfinement transition point: this is at odds with
what is observed around the standard thermal transition, and might
indicate that the transition is stronger in this case.

The localization properties of Dirac modes were studied using the
staggered discretization of the Dirac operator. We found that the
behavior of the low-lying modes at the reconfinement transition is
similar to that observed at the usual confinement phase transition in
the undeformed theory. While the system is in the deconfined phase
in the trivial center sector, the lowest modes of the staggered
Dirac operator are localized both at zero and nonzero deformation
coupling, up to a critical point (mobility edge) in the spectrum. Bulk
modes above the mobility edge are instead delocalized on the whole
lattice. As the deformation coupling grows at fixed compactification
length and the system moves towards the reconfined phase, the mobility
edge decreases, and localized modes eventually disappear as the system
crosses over into the reconfined phase. This is exactly what happens
in standard (undeformed) YM, with the mobility edge decreasing when
the temperature is decreased, and localized modes disappearing when
crossing over to the usual confined phase~\cite{Kovacs:2017uiz}. It is
not clear whether $\lambda_c$ vanishes continuously or discontinuously
at the reconfinement transition: while a discontinuous behavior would
be consistent with the first-order nature of the transition, and with
what we observed for the thermal monopole observables, further studies
are required to make a conclusive statement. In future studies one
could also consider a different discretization of the Dirac operator,
for example the Overlap discretization, as it has been done in
standard YM~\cite{Vig:2020pgq}.

To summarize, we have demonstrated that two well established
phenomena, which are known to characterize the confinement transition
{in $SU(3)$ gauge theory}, characterize the reconfinement transition
as well: condensation of thermal monopoles and delocalization of the
lowest Dirac modes. This indicates that important physical features of
the two transitions are similar, and at the same time establishes a
stronger link between the two phenomena that we have analyzed. It is
reasonable to hypothesize that the presence of thermal monopole
trajectories in a gauge configuration has some reflection on the
eigenmodes of the Dirac spectrum, and that the delocalization of
thermal monopole trajectories, related to the appearance of larger and
larger numbers of wrappings, could be related to the delocalization of
the low-lying Dirac modes. On the other hand, the density of
monopoles seems to have little bearing on it, as seen by contrasting
the usual confinement transition and the reconfining one.  This is
surely something to be further investigated in future studies.

\vspace{0.5cm}
\paragraph*{Acknowledgments}
We thank T.~G.~Kov{\'a}cs and R.~Vig for useful
discussions and for making their results available to us. Numerical
simulations have been performed at the Scientific Computing Center at
INFN-PISA and on the MARCONI machine at CINECA, based on the agreement
between INFN and CINECA (under project INF19\_npqcd, INF20\_npqcd and
IscraB project IsB20\_TDEDGE). MG was partially supported by
  the NKFIH grant KKP-126769.

\end{document}